\documentclass[final, a4paper]{IEEEtran}
\newlength \figwidth
\usepackage{cite}
\ifCLASSINFOpdf
  \usepackage[pdftex]{graphicx}
	\usepackage{epstopdf}
  \graphicspath{{../Figures/}}
  \DeclareGraphicsExtensions{.eps}
\else
  \usepackage[dvips]{graphicx}
\fi
\usepackage[cmex10]{amsmath}
\usepackage[tight,footnotesize]{subfigure}
\usepackage{amssymb}
\usepackage{amsthm}
\usepackage{url}
\usepackage{dsfont}
\usepackage{tabulary}
\usepackage{multirow}
\usepackage{color}

\usepackage{color}
\usepackage{times}
\usepackage{verbatim}
\usepackage{floatflt}
\usepackage{enumerate}
\usepackage{array}
\usepackage{multicol,afterpage,wrapfig} 
\usepackage{tikz}
\usepackage{algorithm}
\usepackage[noend]{algpseudocode}
\makeatletter
\def\BState{\State\hskip-\ALG@thistlm}
\makeatother
\usepackage{amsmath,amssymb,eucal}
\usepackage[utf8]{inputenc}
\usepackage{epsfig}
\usepackage{exscale}
\usepackage{latexsym}
\usepackage{verbatim}
\usepackage{amsfonts}
\usepackage{subfigure}
\usepackage{multirow}
\usepackage{cite}
\usepackage{balance}
\usepackage{color}
\usepackage{transparent}
\usepackage{import}
\usepackage{mathtools}
\usepackage{tikz}
\usetikzlibrary{automata,arrows,positioning,calc}
\usepackage{bbm}
\usepackage[printonlyused,withpage]{acronym}
\usepackage{tabu}
\usepackage{longtable}

\newcommand{\be}{\begin{equation}}
\newcommand{\ee}{\end{equation}}
\newcommand{\ba}{\begin{align}}
\newcommand{\ea}{\end{align}}
\newcommand{\bea}{\begin{eqnarray}}
\newcommand{\eea}{\end{eqnarray}}
\newcommand{\fa}{\mathcal{P}_{\mathrm{fa}} }
\newcommand{\md}{\mathcal{P}_{\mathrm{md}} }
\newcommand{\Tcal}{\mathcal{T}}
\newcommand{\Acal}{\mathcal{A}}
\newcommand{\Pcal}{\mathcal{P}}
\newcommand{\Gcal}{\mathcal{G}}
\newcommand{\Dcal}{\mathcal{D}}
\newcommand{\Pdecl}{\mathcal{P}_{\mathrm{dec},l} }

\newcommand{\Rbk}{R_{b,k} }
\newcommand{\Sbk}{S_{b,k} }
\newcommand{\Ibk}{I_{b,k} }
\newcommand{\Wbk}{W_{b,k} }
\newcommand{\tildeWbk}{\tilde{W}_{b,k} }
\newcommand{\tildeSbk}{\tilde{S}_{b,k} }
\newcommand{\tildeIbk}{\tilde{I}_{b,k} }
\newcommand{\tildeNbk}{\tilde{N}_{b,k} }
\newcommand{\Hone}{\mathcal{H}_1}
\newcommand{\Hzero}{\mathcal{H}_0}
\newcommand{\Es}{E_{\mathrm{s}}}
\newcommand{\Et}{E_{\mathrm{t}}}
\newcommand{\Ed}{E_{\mathrm{d}}}
\newcommand{\Ps}{P_{\mathrm{s}}}
\newcommand{\Pt}{P_{\mathrm{t}}}
\newcommand{\Pd}{P_{\mathrm{d}}}
\newcommand{\Pc}{P_{\mathrm{c}}}
\newcommand{\Ts}{T_{\mathrm{s}}}
\newcommand{\kf}{k_{\mathrm{f}}}
\newcommand{\kd}{k_{\mathrm{d}}}
\newcommand{\kc}{k_{\mathrm{c}}}
\newcommand{\rc}{d_{\mathrm{c}}}

\def\BibTeX{{\rm B\kern-.05em{\sc i\kern-.025em b}\kern-.08em
    T\kern-.1667em\lower.7ex\hbox{E}\kern-.125emX}}

\newcommand*\xbar[1]{%
  \hbox{%
    \vbox{%
      \hrule height 0.5pt 
      \kern0.36ex
      \hbox{%
        \kern-0.12em
        \ensuremath{#1}%
        \kern-0.12em
      }%
    }%
  }%
} 

\newcommand{\indic}{\mathds{1}}

\usetikzlibrary{arrows,calc}
\allowdisplaybreaks

\IEEEoverridecommandlockouts
\begin{document}


\newtheorem{Theorem}{\bf Theorem}
\newtheorem{Corollary}{\bf Corollary}
\newtheorem{Remark}{\bf Remark}
\newtheorem{Lemma}{\bf Lemma}
\newtheorem{Proposition}{\bf Proposition}
\newtheorem{Assumption}{\bf Assumption}
\newtheorem{Definition}{\bf Definition}
\title{Energy Efficiency of Distributed Signal Processing in Wireless Networks: A Cross-Layer Analysis}
\author{Giovanni~Geraci, Matthias~Wildemeersch, and Tony~Q.~S.~Quek
\thanks{G.~Geraci and T.~Quek are with the Singapore University of Technology and Design (e-mail: giovanni\_geraci@sutd.edu.sg, tonyquek@sutd.edu.sg).}
\thanks{M.~Wildemeersch is with the International Institute for Applied Systems Analysis, Laxenburg, Austria (e-mail: wildemee@iiasa.ac.at).}
\thanks{The  material  in  this  paper  has  been  presented in part at the IEEE Global Commun. Conf., San Diego, CA, Dec. 2015 \cite{GerWilQueGLOBECOM15}, and at the 2nd Asia-Pacific Conference on Complex Systems Design \& Management, Singapore, Feb. 2016 \cite{GerWilQueCSDM16}.}
}
\maketitle
\thispagestyle{empty}
\begin{abstract}
In order to meet the growing mobile data demand, future wireless networks will be equipped with a multitude of access points (APs). Besides the important implications for the energy consumption, the trend towards densification requires the development of decentralized and sustainable radio resource management techniques. It is critically important to understand how the distribution of signal processing operations affects the energy efficiency of wireless networks. In this paper, we provide a cross-layer framework to evaluate and compare the energy efficiency of wireless networks under different levels of distribution of the signal processing load: (i) \emph{hybrid}, where the signal processing operations are shared between nodes and APs, (ii) \emph{centralized}, where signal processing is entirely implemented at the APs, and (iii) \emph{fully distributed}, where all operations are performed by the nodes. We find that in practical wireless networks, hybrid signal processing exhibits a significant energy efficiency gain over both centralized and fully distributed approaches.
\end{abstract}
\section{Introduction}

The current growth rate of wireless data exceeds both spectral efficiency improvements and availability of new wireless spectrum, and is therefore driving greater spatial reuse through a larger number of small cells and access points (APs) \cite{AndrewsJSAC2012,QuekBook,LopezPerezSURVEY2015,YanWanGerCOMMSMAG2015}. The trend in cellular networks towards densification and heterogeneity is essential to respond adequately to the continued surge in mobile data traffic. At the same time, the multitude of APs, the heterogeneity of the network architecture, and the density of its topology will make centralized network control impractical and call for a distribution of the signal processing load \cite{tall2014distributed,dimakis2010gossip,DiLorenzoSPMAG2013}. In this article, we aim to evaluate how distributed signal processing affects the performance of wireless networks, and to find those that are most energy efficient.

\subsection{Background and Motivation}

Future wireless networks will not only serve a very dense population of computers, smartphones, and tablets, but will also offer connectivity to a massive number of environmental sensors, control devices, and home appliances \cite{EricssonREPORT2011,CavalcanteSPMAG2014,ChoiTSP2015}. The foreseen increasing number of nodes and traffic will make centralized control and resource management inadequate, and requires the introduction of distributed methods. Distributed control and computation has been well investigated \cite{bamieh2002distributed,bertsekas1989parallel,nedic2009distributed,chen2013distributed}, and has important applications in wireless (sensor) networks in the context of cognitive radio \cite{Cao2012,khajehnouri2007distributed} and self-organizing networks \cite{Aliu2013}. Self-organization, self-configuration, and self-optimization are necessary to manage complexity, to reduce the cost of operation, and to enhance performance and profitability of the network \cite{CombesINFOCOM2012,CombesINFOCOM2013}. Exploiting the cognitive capabilities of both APs and mobile devices is one of the keys to ensure the viability of future wireless networks. The wireless data explosion will break the present network management paradigm and requires the development of distributed radio resource management and signal processing techniques. It is of critical importance to understand how the distribution of signal processing operations will affect the energy efficiency of future wireless networks.


The energy consumption of signal processing operations in wireless networks is contingent on how efficiently the MAC (media access control) layer manages the available resources and determines access opportunities for the nodes. The MAC layer must keep to a minimum those transmissions that are corrupted by interference and therefore jeopardized. The energy efficiency of signal processing also heavily relies on the physical layer, which must be designed to guarantee large throughput while reducing the power consumption \cite{MeshkatiSPMAG2007,YanGerQue:15}. The strong interaction between the MAC and the physical layer in wireless networks calls for a cross-layer design that exploits this interdependency to increase the energy efficiency \cite{Li2011,MiaoBook,XuTSP2014}. Moreover, a cross-layer approach is imperative in order to study the energy efficiency under a distribution of the signal processing load.

\subsection{Approach and Contributions}

The main goal of this paper is to study energy efficiency in wireless networks under different levels of distribution of the signal processing load. We consider the following operations: spectrum sensing in a random topology, media access control, transmission, and multi-user decoding via successive interference cancellation. We explore three scenarios: (i) hybrid, where the signal processing operations are shared between nodes and APs, (ii) centralized, where signal processing is entirely implemented at the APs, and (iii) fully distributed, where all operations are performed by the nodes.\footnote{In the following, we will refer to this scheme as the distributed scheme.} We develop a cross-layer framework to derive the throughput and the energy consumption due to signal processing operations for the whole network, i.e., both nodes and APs. This is a practical problem that has not yet been addressed. In this paper, we consider a network where nodes can be partitioned into clusters, each connected to an AP \cite{GulEvaAndTin_TSP2010,HeathTSP2013}. Depending on the network management approach, transmissions can be centrally scheduled by APs or nodes can access the spectrum via a distributed MAC protocol with spectrum sensing, and colliding transmissions can be resolved at the APs via multi-user decoding (MD).
We provide a general analysis of the MAC protocol that accounts for the interference and for the errors made in the spectrum sensing phase, and we analyze MD by modeling the colliding nodes with a binomial point process (BPP). With the proposed cross-layer framework, we can explicitly characterize the energy consumption due to sensing, control, transmission, and decoding operations, as well as the throughput and ultimately the energy efficiency of the network. Our main contributions are summarized below.
\begin{itemize}
\item We provide a cross-layer framework to assess the energy efficiency of wireless networks under hybrid, centralized, and distributed signal processing load. Our framework accounts for spectrum sensing, network access, and decoding performed at nodes and APs.
\item We derive the probabilities of missed detection and false alarm of an energy-detection-based spectrum sensing scheme in a random clustered topology. We quantify how these probabilities affect the throughput and energy consumption of a random distributed MAC protocol.
\item We analyze the performance of multi-user decoding via successive interference cancellation in a BPP of colliding nodes, and we make the relation between the probability of successful decoding and the transmission rate explicit.
\item We compare the energy efficiency under different levels of distribution of the signal processing load. We find that in practical wireless networks, hybrid signal processing exhibit a significant energy efficiency gain over both centralized and fully distributed approaches. 
\end{itemize}

The remainder of the paper is organized as follows. The system model is introduced in Section II. In Section III, we derive the probabilities of missed detection and false alarm of a spectrum sensing scheme in a random topology. In Section IV, we obtain the energy consumption of a random MAC protocol with imperfect sensing. In Section V, we analyze the performance of MD via successive interference cancellation. In Section VI, we compare the energy efficiency of hybrid signal processing to fully centralized/distributed approaches. The paper is concluded in Section VII.
\section{System Model} 

\subsection{Topology and Access Scheme}

We consider the uplink of a wireless network where nodes can be partitioned into groups, or clusters. We assume that each cluster has an access point, and that each node in the cluster is randomly placed in the neighborhood of the AP \cite{HongMobilityModel}. Our model is general and can capture various network architectures such as heterogeneous networks, ad hoc networks, etc. \cite{XuTSP2014,DongTSP2008,WebAndJin_TIT2007}. The locations of all nodes in the cluster are uniformly distributed according to a Poisson point process (PPP) of density $\lambda$ in a circular area of radius $\rc$ and centered in $x$, represented by $b(x, \rc)$, with $M=\lambda \pi \rc^2$ the average number of nodes in each cluster.\footnote{Our model naturally captures ad hoc networks, and it is general enough to capture the uplink of a cellular network. In fact, we can reproduce the results in \cite{Novlan2013} by adjusting the parameter $d_\mathrm{min}$ introduced in Assumption \ref{approx:cluster_approx}.} Let $\rc$ be the cluster radius and let $x$ be the location of the AP. For ease of notation, we use $x_{\mathrm{h},i}$ to indicate the $i$-th AP, as well as its location. We will refer to the cluster centered around the origin as the representative cluster, and nodes located outside this cluster contribute to the interference. Outside the representative cluster $b(0, \rc)$, the parent process of APs $x_{\mathrm{h},i}$ follows a PPP with density $\lambda_\mathrm{h}$. Since the active nodes are uniformly distributed within the coverage area $b(x_{\mathrm{h},i}, \rc)$ of the AP $x_{\mathrm{h},i}$, the total set of interfering nodes in uplink forms a Matern cluster process denoted by $\Psi$ \cite{GanHae_TIT2009}.

Each AP receives messages from all nodes in the uplink. We assume that the nodes use a strategy based on orthogonal frequency channels, where the available bandwidth is partitioned into a set of $N$ multiple closely spaced subcarriers.\footnote{Our results are general and hold under different multiple access schemes. In this respect, frequency division, time division, and orthogonal code division are equivalent as they all divide the spectrum orthogonally \cite{GuidoCL2014}.} Nodes use subsets of subcarriers, and this allows simultaneous data transmission from several nodes. Network management is then achieved by means of a \emph{hybrid} signal processing scheme, where the nodes employ a MAC protocol that builds on a spectrum sensing functionality, and the APs employ multi-user decoding to resolve collisions arising from the random access protocol.

\subsection{Channel Model}

We consider single-antenna nodes,\footnote{Our analysis can be extended by considering multi-antenna access points that employ spatial multiplexing \cite{FerGerQueWin_1:2015}.} and the channels between any pairs of nodes are assumed to be independent and identically distributed (i.i.d.) and quasi-static, i.e., constant during the transmission of a frame. We assume that each channel is narrowband and affected by two attenuation components, namely path loss and fading.\footnote{Although the presence of a line-of-sight component is likely within clusters, the analysis presented here is based on Rayleigh fading for reasons of tractability. Note that the results involving the machinery of stochastic geometry can be adjusted for an arbitrary fading distribution building on stochastic equivalence and a scaling of the node densities \cite{BlaKee_PIMRC13}.} Let $a$ be a random node located in cluster $\mathcal{I}$. The received signal at the random node $a$ can be written as
\begin{equation}
r(t) = s(t) + i(t) + w(t)
\label{eqn:channel_model}
\end{equation}
where $s(t)$ is the signal received at node $a$ from other nodes in the same cluster $\mathcal{I}$, given by
\begin{equation}
s(t) = 	\sum_{j \in \mathcal{I} \backslash a} d_{j}^{-\frac{\alpha}{2}} h_{j} u_j(t),
\end{equation}
whereas $i(t)$ is the interference received from other clusters, given by
\begin{equation}
i(t) = 	\sum_{j \in \Psi \backslash \mathcal{I}} d_{j}^{-\frac{\alpha}{2}} h_{j} u_j(t),
\end{equation}
and where $\alpha$ is the path loss exponent, $u_j(t)$ is the signal transmitted by node $j$, $d_{j}$ and $h_{j}\sim\mathcal{CN}(0,1)$ are the distance and the Rayleigh fading coefficient between nodes $a$ and $j$, respectively, and $w(t) \sim \mathcal{CN}(0, \sigma^2_{\mathrm{w}})$ is additive complex white Gaussian noise.

\subsection{Energy Efficiency}

Under a hybrid signal processing scheme, we can identify three main contributions to the energy consumption of the wireless network, namely (i) the sensing energy at all nodes, (ii) the transmission energy at all nodes, and (iii) the decoding energy at the APs. We consider the energy consumption of the entire network, therefore energy-efficiency tradeoffs will be such that the savings at the APs are not counteracted by increased consumption at the nodes, and vice versa \cite{Feng2013}. The energy consumption in each cluster per subcarrier and per time slot can be modeled as
\begin{equation}
E = \Es + \Et + \Ed
\end{equation}
where $\Es$, $\Et$, and $\Ed$ are the energy consumption due to sensing, transmission, and decoding, respectively. For each node that senses the spectrum occupation, the corresponding sensing energy consumption is proportional to the sensing power $\Ps$ and to the sensing time $\Ts$. Similarly, the transmission energy $\Et$ of a node is proportional to the transmit power $\Pt$ and to the total transmission time of the node. The decoding energy consumption $\Ed$ is incurred at the AP during the decoding process, and it is assumed proportional to the decoding power $\Pd$, to the time slot duration $T$, and to the total number of decoding attempts.\footnote{We neglect the dependence of $\Pt$ and $\Pd$ on the modulation used \cite{Cui2005}.}

We denote by $\chi(\zeta)[\frac{\textrm{bits}}{\textrm{s}}]$ a spectral gain that accounts for the modulation scheme used and for the bandwidth of each subcarrier, where $\zeta$ is the SINR (signal-to-interference-plus-noise ratio) decoding threshold. The throughput $R$ of the wireless network is defined as the mean number of bits successfully transmitted to each AP per subcarrier and per time slot. Finally, the energy efficiency $\eta=\frac{R}{E}$ is defined as the number of bits successfully transmitted per joule of energy spent \cite{Feng2013}.

\begin{table}
\centering
\caption{Notation Summary}
\label{table:notationtable}
\begin{tabulary}{\columnwidth}{ | c | C | }
\hline
    \textbf{Notation} & \textbf{Description} \\ \hline
		$\eta$; $R$; $E$ 		&	Energy efficiency, throughput, and energy consumption with a hybrid scheme  \\ \hline
		$\Es$; $\Et$; $\Ed$; $\chi$; $\zeta$		&	Sensing, transmission, and decoding energy; spectral gain; decoding threshold  \\ \hline
		$\Ps$; $\Pt$; $\Pd$; $\Pc$		&	Sensing, transmission, decoding, and control channel power per subcarrier \\ \hline
	  $\rc$; $\lambda_{\mathrm{h}}$;	$\lambda$; $M$	&	Cluster radius; density of APs; density of nodes; mean number of nodes per cluster  \\ \hline
		$\alpha$; $h$; $d_j$	&	Path loss exponent; fading coefficient; distance between a given node and node $j$   \\ \hline
		$\fa$; $\md$; $\Ts$; $B$		&	Prob. of false alarm; prob. of missed detection; sensing time; sensing blocks   \\ \hline
    $E_k$; $\rho$;	$\hat{\mathbf{q}}$	&	Received energy on subcarrier $k$; sensing threshold; estimated spectrum occupancy  \\ \hline
		$\Ibk$; $\sigma_{\mathrm{I}}^2$; $d_\mathrm{min}$		&	Inter-cluster interference; variance of $\Ibk$; minimum distance from the interferers  \\ \hline
		$T$; $\kf$; $\kc$; $\kd$		&	Slot duration; number of slots in a frame; contention slots; contention-free slots  \\ \hline
		$p$; $N$; $N_{\mathrm{f},t}$; $M_{\mathrm{i},t}$ 		&	Spectrum access prob.; number of subcarriers; free subcarriers at $t$; inactive nodes \\ \hline
		$\mathbf{S}_l$; $\Pcal_{l,t}$; $\Tcal_{i,l,t}$		&	State with $l$ nodes on a subcarrier; prob. of $\mathbf{S}_l$; transition prob. from $\mathbf{S}_i$ to $\mathbf{S}_l$  \\ \hline
		$s$; $\mu_t$ 		&	 Maximum number of subcarriers per node; mean number of collisions  \\ \hline
		$\Dcal_{i,l}$; $\Pdecl(n)$		&	Prob. decoding $i$ out of $l$ transmissions; prob. decoding the $n$-th strongest out of $l$  \\ \hline
		$\eta_{\mathrm{C}}$; $R_{\mathrm{C}}$; $E_{\mathrm{C}}$		&	Energy efficiency, throughput, and energy consumption with a centralized scheme  \\ \hline
		$\eta_{\mathrm{D}}$; $R_{\mathrm{D}}$; $E_{\mathrm{D}}$		&	Energy efficiency, throughput, and energy consumption with a distributed scheme  \\ \hline
\end{tabulary}
\end{table}
\section{Analysis of Spectrum Sensing}

In this section, we analyze the performance of a spectrum sensing scheme by deriving the probabilities of missed detection and false alarm. Spectrum sensing is used by each node to obtain information on the local subcarrier occupancy, and the probabilities of missed detection and false alarm affect the performance of the MAC protocol and the energy efficiency of the network, as will be discussed in Section IV.

\subsection{Preliminaries}

In a hybrid signal processing scheme, spectrum sensing is implemented at each node to reliably detect the transmissions occurring in its cluster with a low probability of false alarm $\fa$ (to guarantee high spectral utilization) and a low probability of missed detection $\md$ (to guarantee a small number of colliding transmissions). In the following, we assume that each node in the network applies spectrum sensing by means of an energy detector (ED). Although other detection schemes have been proposed in the literature \cite{tandra_JSTSP2008,BhaMur_SPAWC2010}, the ED scheme is particularly attractive and widely used since it incurs low computational complexity and low power consumption \cite{VanTrees:Wiley01,axell_IEEESPM2012}. We note that the analysis provided in the following sections holds under different spectrum sensing schemes by simply replacing the values of $\fa$ and $\md$.

If we denote by $q_k$ the occupancy status of subcarrier $k$ within the cluster, the spectrum sensing problem can be regarded as the decision process of whether the subcarrier $k$ is vacant, i.e., $q_k = 0$, or occupied, i.e., $q_k = 1$. We denote by $\hat{\mathbf{q}} = [\hat{q}_1,\ldots,\hat{q}_N]$ the estimated spectrum occupancy vector at a given sensing node, and by $\Ts$ the total sensing time. The time interval $\Ts$ must be small compared to the channel coherence time, such that the spectrum occupancy is block stationary. During the spectrum sensing interval, each node samples the received signal $r(t)$ at the Nyquist rate $\mathsf{R_N}$, obtaining the sequence
\begin{equation}
r_n = r\left(n/\mathsf{R_N}\right), \quad n=1,\ldots,T_{s}\mathsf{R_N}.
\end{equation}
The sequence $r_n$ is then divided into $B$ blocks of $N$ samples, with $N$ corresponding to the number of subcarriers, such that the total sensing time is given by $\Ts=\frac{B N}{\mathsf{R_N}}$. The $b$-th block, $b=1,\ldots,B$, can be represented by its $N$-point discrete Fourier transform (DFT)
\begin{equation}
\Rbk \!=\!  \frac{1}{\sqrt{N}} \! \sum_{n=(b-1)N}^{bN-1} \!{r_n e^{-j2\pi \frac{nk}{N}}}, \enspace k=1,\ldots,N.
\end{equation}
The samples $|\Rbk|^2$ contain the energy received by the sensing node on subcarrier $k$ in the $b$-th block. For each subcarrier, the node computes the summary statistics as the average received signal energy over the $B$ blocks, given by
\begin{equation}
E_{k} = \frac{1}{B} \sum_{b=1}^{B}{\left| \Rbk \right|^2}, \quad k=1,\ldots,N
\label{eqn:Ek}
\end{equation}
then obtaining the estimated spectrum occupancy \cite{Quan2009WidebandSS}
\begin{equation}
\hat{q}_k = \indic_{(E_{k} > \rho)}, \enspace k=1,\ldots,N, 	
\end{equation}
where $\indic_{(\cdot)}$ is the indicator function. The choice of the decision threshold $\rho$ should be a tradeoff between the probabilities of false alarm $\fa$ and missed detection $\md$, since increasing $\rho$ yields a smaller $\fa$ and a larger $\md$, and vice versa \cite{VanTrees:Wiley01}.

\subsection{Missed Detection and False Alarm}

We now analyze the performance of the spectrum sensing scheme by deriving the probabilities of missed detection and false alarm \cite{WilQueRabSlu_TC2013}. The spectrum occupancy estimation is a binary hypothesis test problem for each subcarrier. The two hypotheses $\mathcal{H}_1$ and $\mathcal{H}_0$ correspond to the cases when the subcarrier is being used or not being used by one or more nodes in the same cluster, respectively. This is consistent with the multi-user decoding scheme analyzed in Section V, where concurrent transmissions from other clusters are treated as interference, whereas concurrent transmissions within a cluster are regarded as collisions and can be resolved by the AP.

Let $\Sbk$, $\Ibk$, and $\Wbk$ be the $N$-point DFTs of $s(t)$, $i(t)$, and $w(t)$, respectively, over the $b$-th block. The DFT of the signal received at the typical sensing node on subcarrier $k$ over the $b$-th block under the two hypotheses above can be written as
\begin{equation}
\begin{aligned}
\mathcal{H}_0: & \enspace \Rbk = \Ibk + \Wbk \\
\mathcal{H}_1: & \enspace \Rbk = \Sbk + \Ibk + \Wbk.
\end{aligned}
\end{equation}
The probability of missed detection is defined as the probability that the decision variable $E_k$ falls under the threshold $\rho$ under hypothesis $\mathcal{H}_1$, and it is given by
\begin{equation}
\md = \textrm{Pr}[E_k < \rho | \mathcal{H}_1].
\end{equation}
The probability of false alarm is defined as the probability that $E_k$ surpasses the threshold under hypothesis $\mathcal{H}_0$, and it is given by
\begin{equation}
\fa = \textrm{Pr}[E_k > \rho | \mathcal{H}_0].
\end{equation}

Let $\mu$ be the average number of colliding nodes per cluster on a given subcarrier, which depends on the MAC protocol and will be derived in (\ref{eqn:mu}) as a function of the time slot $t$, and let us assume that colliding nodes are uniformly distributed within each cluster.\footnote{In a system where most collisions are caused by missed detection events, nodes close to each other are less likely to transmit simultaneously. However, as will be shown in Fig. \ref{fig:ROC}, $\md$ is typically small, which implies that most collisions are due to the randomness of the MAC protocol \cite{GeraciICC15} and are therefore location independent.} We now make the following approximation.

\begin{Assumption}
We approximate the Matern cluster process $\Psi$ of the interfering nodes by a PPP $\Phi$ with density $\mu \lambda_\mathrm{h}$. We neglect the location-dependence and assume a constant exclusion distance $d_\mathrm{min}$ between the sensing node and the closest out-of-cluster interferer. As a result, the amplitude of the aggregate network interference can be expressed as 
\be 
\Ibk =\sum_{j \in \Phi \backslash b(0, d_\mathrm{min})} \sqrt{\Pt} |h_j| d_j^{-\alpha/2} \, ,
\label{eq.aggIntAmp}
\ee
where $h_j$ is the fading coefficient between node $j$ and the sensing node, and $\Pt$ is the transmission power relative to subcarrier $k$.
\label{approx:cluster_approx}
\end{Assumption}

\begin{Remark}
We note that from the displacement theorem \cite[Theorem 1.10]{BacBla_NOW2009}, the PPP approximation is exact for $\mu=1$, which is a practical value under well-designed MAC protocols that avoid collisions. Moreover, we note that assuming an exclusion region between the sensing node and the closest out-of-cell interferer is equivalent to considering non-overlapping clusters, which is practically more relevant. The accuracy of Assumption \ref{approx:cluster_approx} will be validated in Fig. \ref{fig:cluster_approx}. 
\label{remark:assumptions}
\end{Remark}

The exclusion region in (\ref{eq.aggIntAmp}) leads to a bounded path loss model where the distribution of the aggregate interference $\Ibk$ has finite moments \cite{WinPinShe_ProcIEEE2009, RabQueShiWin_JSAC2011}. Therefore, building on the central limit theorem, this allows the following approximation \cite{Inaltekin2012gaussian}.

\begin{Assumption}
We use a Gaussian distribution to model the aggregate interference as
\be
\Ibk \sim \mathcal{N}(\mu_{\mathrm{I}}, \sigma_{\mathrm{I}}^2)	\, ,
\ee
where the moments $\mu_{\mathrm{I}}$ and $\sigma_{\mathrm{I}}^2$ are derived in the following lemma.
\end{Assumption}

\begin{Lemma}
The moments of the aggregate interference $\Ibk$, modeled as a Gaussian-distributed 
random variable, are given by $\mu_{\mathrm{I}} = 0$ and
\be
\sigma^2_{\mathrm{I}} = \Pt \frac{\pi \mu \lambda_\mathrm{h}}{2\alpha -1} d_\mathrm{min}^{2-\alpha} \mu_{|h|,2} \, ,
\label{eqn:moments2}
\ee
where $\mu_{|h|,2}$ represents the second moment of the fading distribution.
\label{Lemma:moments}
\end{Lemma}
\begin{IEEEproof}
See Appendix \ref{App:moments}.
\end{IEEEproof}

We now obtain the probabilities of missed detection and false alarm for the spectrum sensing scheme.

\begin{Lemma}
The probabilities of missed detection $\md$ and false alarm $\fa$ are given by
\begin{multline}
\md = \mathrm{Pr}[E_k < \rho | \Hone] = \\  
\frac{1}{2} + \frac{1}{2\pi}\!\int_{0}^{\infty}\!\!\!\mathfrak{Re}\!\left\{ \frac{\psi_{E_k|\Hone}(-j\omega)e^{j \omega\rho} \!-\!\psi_{E_k|\Hone}(j\omega)e^{-j \omega \rho}}{j \omega} \right\} d\omega
\label{eqn:md} 
\end{multline}
and
\begin{multline}
\fa = \mathrm{Pr}[E_k > \rho | \Hzero] = \\  
\frac{1}{2} - \frac{1}{2\pi}\!\int_{0}^{\infty}\!\!\!\mathfrak{Re}\!\left\{ \frac{\psi_{E_k|\Hzero}(-j\omega)e^{j \omega\rho} \!-\!\psi_{E_k|\Hzero}(j\omega)e^{-j \omega \rho}}{j \omega} \right\} d\omega ,
\label{eqn:fa}
\end{multline}
where $\psi_{E_k|\Hone}(j\omega)$ and $\psi_{E_k|\Hzero}(j\omega)$ represent the characteristic function (CF) of $E_k$ under hypotheses $\Hone$ and $\Hzero$, respectively, given by
\be
\psi_{E_k|\Hone}(j\omega) = \frac{\left(  \mathbb{E}_{d_i} \left[ \frac{1}{1 - j\omega(\Pt/(2 d_i^{\alpha}) + \sigma_{\mathrm{IN}}^2)} \right] \right)^l}{\left( 1-2j\omega\sigma_{\mathrm{IN}}^2 \right)^{B/2-1}} ,
\ee
\begin{align}
\psi_{{E_k}|\Hzero}(j\omega) = \frac{1}{\left(1-2j\omega\sigma_{\mathrm{IN}}^2 \right)^{B/2}},
\end{align}
where $\sigma^2_\mathrm{IN} = \frac{\sigma_{\mathrm{I}}^2+\sigma_{\mathrm{w}}^2}{B}$, $l$ is the number of active nodes on subcarrier $k$ in the representative cluster, and $d_i$ is the distance between the typical sensing node and any other node $i$ in the same cluster, with probability density function (pdf) given by
\be
f_{d_i}(x) = \frac{2x}{\rc^2} \left( \frac{2}{\pi} \cos^{-1}\left(\frac{x}{2 \rc}\right) - \frac{x}{\pi \rc} \sqrt{1-\frac{x^2}{4 \rc^2}} \right).
\label{eqn:pdf_ri}
\ee
\label{Lemma:PmdPfa}
\end{Lemma}
\begin{IEEEproof}
See Appendix \ref{App:PmdPfa}.
\end{IEEEproof}

\subsection{Validation and Insights}

We now provide numerical results to confirm the accuracy of the assumptions made in this section and to show the performance of the spectrum sensing scheme. The probabilities of missed detection and false alarm affect the performance of the MAC protocol and therefore the energy efficiency of the network. The exact relation between sensing performance and energy consumption will be made explicit in Section IV.

\begin{figure}[!t]
\centering
\includegraphics[width=\figwidth]{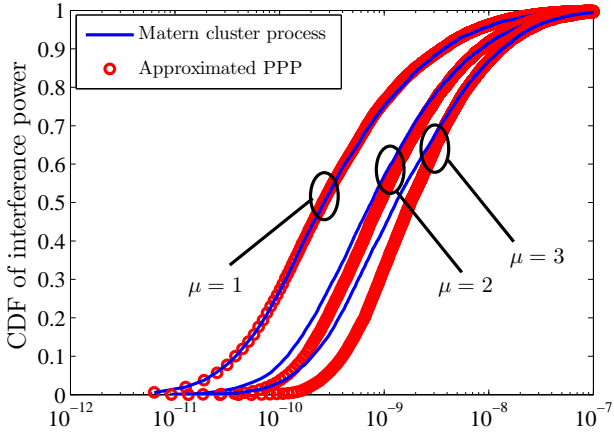}
\caption{Comparison between the simulated cumulative distribution functions (CDFs) of the interference obtained as a Matern cluster process and as an approximated PPP, respectively, for an average of $\mu=1$, $2$, and $3$ concurrent transmissions per cluster, $\rc=100$, and $d_{\mathrm{min}}=\rc$.}
\label{fig:cluster_approx}
\end{figure}

In Fig. \ref{fig:cluster_approx}, we compare the simulated cumulative distribution functions (CDFs) of the interference power obtained as a Matern cluster process and as an approximated PPP, respectively. Figure \ref{fig:cluster_approx} shows a perfect match for $\mu=1$ active node per subcarrier per cluster, which is a practical value under well-designed MAC protocols. On the other hand, the accuracy degrades for higher and less practical values of $\mu$, when the PPP approximation tends to be conservative and slightly overestimates the interference distribution. This confirms the accuracy of the approximation proposed in Assumption \ref{approx:cluster_approx} as well as the claims made in Remark~\ref{remark:assumptions}.

\begin{figure}[!t]
\centering
\includegraphics[width=\figwidth]{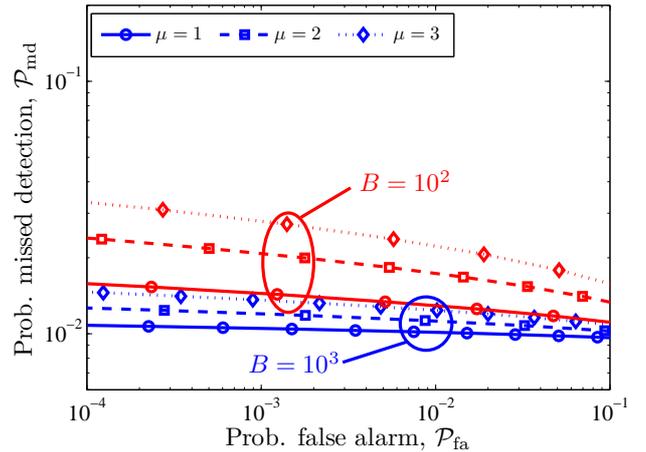}
\caption{Probability of missed detection $\md$ vs false alarm $\fa$ for $l=1$ active user in the representative cluster, an average of $\mu=1$, $2$, and $3$ concurrent transmissions in the other clusters, $B=10^2$ and $10^3$ sensing blocks, $\rc=100$, and $d_{\mathrm{min}}=\rc$.}
\label{fig:ROC}
\end{figure}

In Fig. \ref{fig:ROC}, we illustrate the detection capability of the spectrum sensing scheme by means of the receiver operating characteristic (ROC). Our metrics of interest are the probabilities of missed detection $\md$ and false alarm $\fa$, which affect the performance of the MAC protocol, as will be discussed in Section IV. We note that a small value of $\md$ is especially desirable since missed detection may lead to colliding transmissions on the same subcarrier \cite{zhao_SPM2007}. The ROC curve in Fig. \ref{fig:ROC} shows the tradeoff between $\md$ and $\fa$ by varying the decision threshold $\rho$. Note that the proposed framework is able to quantify the improvement of the detection performance by increasing the number of sensing blocks $B$ for a scenario with random topology. Figure \ref{fig:ROC} shows that with a sufficient number of sensing blocks $B$, the ED-based spectrum sensing scheme can achieve probabilities of missed detection and false alarm of the order of $10^{-2}$.

\section{Analysis of Media Access Control}

In this section, we analyze the energy consumption of all nodes in a cluster due to sensing and transmission when a MAC protocol is employed to access the spectrum in a distributed manner. In addition, we provide simulations that confirm the accuracy of our analysis. In order to maintain tractability, in the following we consider a stylized MAC protocol which captures all the key features of distributed random access schemes, as discussed in the sequel. We note that our proposed cross-layer framework holds under more general conditions and applies to different MAC protocols by simply replacing the statistics of the number of nodes that occupy a given subcarrier at a certain time slot. A thorough analysis of standard network access protocols, e.g., WLAN (IEEE 802.11) and WPAN (IEEE 802.15), is beyond the scope of this work and can be found, among others, in \cite{Tinnirello2010,Rashwand2015} and references therein.

\subsection{Preliminaries}

In a hybrid signal processing scheme, a random access MAC protocol is implemented at all nodes, who independently attempt to occupy the subcarriers when they are sensed free. The nodes obtain the local channel activity information on all subcarriers via a spectrum sensing scheme, as discussed in Section III. Using random spectrum access may lead to colliding transmissions, which occur if two or more nodes simultaneously start using a subcarrier they sensed as free, or if a node cannot sense the transmission of another node due to the missed detection events analyzed in Section III. On the other hand, random access exhibits several advantages over scheduled access, since  it does not require a control channel, it relieves APs from any centralized scheduling burden, and it does not require feedback overhead from the nodes nor their cooperation \cite{TsatsanisTSP2000,zhao_SPM2007}.

In this section, we consider a random access protocol where each time frame is divided into (i) a slotted contention period when both sensing and transmission can be performed and (ii) a contention-free period reserved for data transmission only.\footnote{At this stage we assume that synchronization is perfectly achieved. The impact of synchronization errors could be object of future research.} We denote by $\kf$ the total number of slots in a frame, and by $\kc$ and $\kd$ the number of contention and contention-free slots, respectively, with $\kf=\kc+\kd$. At the beginning of each contention slot, each node starts sensing the spectrum with probability $p$, by using the spectrum sensing scheme as discussed in Section III, and thus obtains the spectrum occupancy estimation $\hat{\mathbf{q}}$. The node has then two options: if no subcarriers are sensed as locally free, the node defers transmission until the next frame, whereas if $\|\hat{\mathbf{q}}\|_1>0$ subcarriers are sensed as free, the node randomly chooses $\breve{s}$ of them, where $\breve{s}=\min(s,\|\hat{\mathbf{q}}\|_1)$ and $s$ is the maximum number of subcarriers that each node is allowed to use, and it transmits on the selected subcarriers until the end of the frame.

The MAC protocol considered in this section has the following features: (i) spectrum sensing is performed at most once in a time frame, therefore reducing the sensing energy consumption, and (ii) each node randomly selects some of the available subcarriers, therefore collisions only last for a time frame or less.

\subsection{Energy Consumption}

We now analyze the sensing and transmission energy consumption at all nodes when a random access MAC protocol is used under hybrid signal processing. 

The probabilities of missed detection $\md$ and false alarm $\fa$ derived in Section III depend on the number of colliding nodes per cluster, which varies across time slots. However, as shown in Fig. \ref{fig:ROC}, $\md$ and $\fa$ are typically small and therefore do not significantly affect the behavior of the MAC protocol \cite{GeraciICC15}. We can then approximate $\md$ and $\fa$ with constant values chosen as upper bounds on the quantities (\ref{eqn:md}) and (\ref{eqn:fa}), thus providing conservative bounds on the performance of the MAC protocol.\footnote{Such upper bounds can be obtained by noting from Fig. \ref{fig:ROC} that both $\md$ and $\fa$ increase with the number of colliding transmissions $\mu$, and that under practical well-designed MAC protocols, the value of $\mu$ must be kept close to one, for example by adjusting the parameters $p$ and $s$.}

\begin{figure}[t]
\centering
\begin{tikzpicture}[->, >=stealth', auto, semithick, node distance=2.5cm]
\tikzstyle{every state}=[fill=white,draw=black,thick,text=black,scale=1]
\node[state]    (S0)                     {$\mathbf{S}_0$};
\node[state]    (S1)[right of=S0]   {$\mathbf{S}_1$};
\node[state]    (S2)[below of=S1]   {$\mathbf{S}_2$};
\node[state]    (S3)[below of=S0]   {$\mathbf{S}_3$};
\path
(S0) edge[loop left]     node{$\Tcal_{0,0,t}$}         (S0)
    edge[above]     node{$\Tcal_{0,1,t}$}     (S1)
    edge[bend left,left]      node{$\Tcal_{0,2,t}$}      (S2)
    edge[left]    node{$\Tcal_{0,3,t}$}      (S3)
(S1) edge[loop right]     node{$\Tcal_{1,1,t}$}         (S1)
    edge[right]     node{$\Tcal_{1,2,t}$}     (S2)
    edge[bend left,left]      node{$\Tcal_{1,3,t}$}      (S3)
(S2) edge[loop right]     node{$\Tcal_{2,2,t}$}         (S2)
    edge[below]     node{$\Tcal_{2,3,t}$}     (S3)
(S3) edge[loop left]     node{$\Tcal_{3,3,t}$}         (S3);
\end{tikzpicture}
\caption{Transition probabilities $\Tcal_{i,l,t}$ between the various states $\mathbf{S}_l$ for a cluster with $M=3$ nodes.}
\label{fig:Markov}
\end{figure}
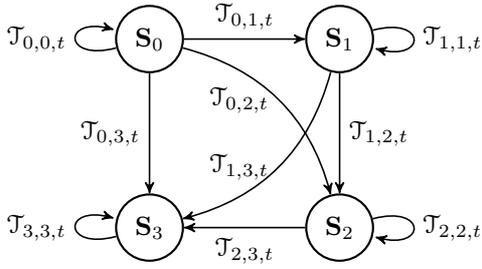

For a given subcarrier, we denote $\mathbf{S}_l$ the state where the subcarrier is occupied by $l$ nodes. At time slot $t$, the probability of the subcarrier being in state $\mathbf{S}_l$ is denoted $\Pcal_{l,t}$, with $\Pcal_{0,1}=1$ since all subcarriers are free at the beginning of the frame. As illustrated in Fig. \ref{fig:Markov}, the probability that a certain subcarrier will be in state $\mathbf{S}_l$ at time slot $t$, given that it is in state $\mathbf{S}_i$ at time slot $t-1$, is denoted by the transition probability $\Tcal_{i,l,t}$, where $\Tcal_{i,l,t} = 0$ $\forall t$ if $i>l$. The resulting Markov chain will be of use not for the study of the stationary distribution, but rather to characterize the transient behavior of the expected carrier occupancy.

We define as inactive nodes those nodes that have not yet sensed the spectrum, and as free subcarriers those subcarriers that are not being occupied within the cluster. We now use the following approximation for the number of inactive nodes and free subcarriers, which will be validated via simulations in Section IV-C.
\begin{Assumption}
We approximate the number of inactive nodes, the number of free subcarriers, and the number of subcarriers sensed as free by a given node at the beginning of time slot $t$ with their respective average values $M_{\mathrm{i},t}$, $N_{\mathrm{f},t}$, and $\hat{N}_{\mathrm{f},t}$.
\label{Ass:mean_field}
\end{Assumption}

By taking into account that each node that activates randomly chooses $\min(s,\hat{N}_{\mathrm{f},t})$ subcarriers, and by defining $\xi_t \triangleq \min(s/\hat{N}_{\mathrm{f},t},1)$, we obtain
\begin{equation}
M_{\mathrm{i},t} = M_{\mathrm{i},t-1} \left(1-p\right) = M_{\mathrm{i},1} \left(1-p\right)^{t-1}, \enspace  t \leq \kc+1
\label{eqn:M}
\end{equation}
\begin{equation}
N_{\mathrm{f},t} = N_{\mathrm{f},t-1} \left(1 - p \left(1-\fa\right) \xi_{t-1} \right)^{M_{\mathrm{i},t-1}}, \enspace  t \leq \kc+1
\label{eqn:N}
\end{equation}
\begin{equation}
\hat{N}_{\mathrm{f},t} = N_{\mathrm{f},t} \left(1-\fa\right) + \left(N_{\mathrm{f},1}-N_{\mathrm{f},t}\right) \md, \enspace  t \leq \kc
\end{equation}
where $M_{\mathrm{i},1}=M$ and $N_{\mathrm{f},1}=N$ at the beginning of the time frame. The equations above account for the fact that nodes can start concurrent transmissions on occupied subcarriers (due to missed detection events) and that a free subcarrier can be sensed as occupied and therefore ignored (due to false alarm events).
We now derive the energy consumption due to spectrum sensing.

\begin{Lemma}
The sensing energy consumption $\Es$ per subcarrier incurred by all nodes in a cluster during a time slot is given by
\begin{equation}
\Es = \frac{\Ps \Ts M}{\kf} \left[1 - (1-p)^{\kc}\right].	
\label{eqn:Es}
\end{equation}
\label{Lemma:Es}
\end{Lemma}
\begin{IEEEproof}
Equation (\ref{eqn:Es}) follows from (\ref{eqn:M}), by noting that $M-M_{\mathrm{i},\kc+1}$ is the mean number of nodes that activate to perform spectrum sensing during a time frame, and by dividing by the number of slots $\kf$ in a frame.
\end{IEEEproof}

In the following, we approximate by $\Acal_{i,t}$ the probability that $i$ nodes activate at time slot $t$ and by $\Gcal_{i,j,t}$ the probability that $j$ nodes choose a certain free subcarrier if $i$ nodes have activated, given by
\begin{equation}
\Acal_{i,t} = \binom{\bar{M}_{\mathrm{i},t}}{i} p^i \left(1-p\right)^{\bar{M}_{{\mathrm{i},t}}-i}
\end{equation}
and
\begin{equation}
\Gcal_{i,j,t} = \binom{i}{j} \left(\left(1-\fa\right)\xi_{t}\right)^j \left(1-\left(1-\fa\right)\xi_{t}\right)^{i-j},
\end{equation}
respectively, where $\bar{M}_{\mathrm{i},t}$ is the nearest integer to $M_{\mathrm{i},t}$. The probability that a subcarrier is in a state $\mathbf{S}_l$ at a certain time slot is then obtained as follows.

\begin{Lemma}
The probability $\Pcal_{l,t}$ that a subcarrier is in state $\mathbf{S}_l$ at time slot $t$ is given by
\begin{equation}
\Pcal_{l,t} = \sum_{i=0}^{l} \Pcal_{i,t-1} \Tcal_{i,l,t}, \enspace l=1,\ldots,M
\label{eqn:P}
\end{equation}
where $\Pcal_{0,1}=1$ and the transition probabilities $\Tcal_{i,l,t}$, $i \leq l$, $t=1,\ldots,\kc$, are given by
\begin{equation}
\Tcal_{i,l,t} = \binom{\bar{M}_{\mathrm{i},t}}{l-i}  \left(p \md \xi_{t}\right)^{l-i} 
 \left(1-p \md \xi_{t}\right)^{\bar{M}_{\mathrm{i},t}-l+i}, \enspace i,l>0
\label{eqn:Til}
\end{equation}
\begin{equation}
\Tcal_{0,l,t} = \sum_{i=l}^{\bar{M}_{\mathrm{i},t}} \Acal_{i,t} \Gcal_{i,l,t}, \enspace l>0 \label{eqn:T0l}
\end{equation}
\begin{equation}
\Tcal_{0,0,t} = \left(1 - p \left(1-\fa\right) \xi_{t}\right)^{\bar{M}_{\mathrm{i},t}}
\label{eqn:T00}.
\end{equation}
\label{Lemma:P}
\end{Lemma}
\begin{IEEEproof}
See Appendix \ref{App:P}.
\end{IEEEproof}

\begin{Corollary}
The mean number of nodes $\mu_t$ that occupy a subcarrier at time slot $t$ can be obtained as
\begin{equation}
\mu_t = \sum_{l=1}^{M} l \Pcal_{l,t}.
\label{eqn:mu}
\end{equation}
\label{corollary:mu}
\end{Corollary}
\begin{IEEEproof}
Equation (\ref{eqn:mu}) follows by averaging the state probabilities $\Pcal_{l,t}$ in Lemma \ref{Lemma:P}.
\end{IEEEproof}

By approximating $\xi_{t} \approx \frac{s}{N_{\mathrm{f},t}}$, by assuming a small probability of false alarm $\fa \ll 1$, and by using Taylor series expansion, the average number $N_{\mathrm{f},t}$ of free subcarriers at time slot $t$ as expressed in \eqref{eqn:N} can be approximated as
\begin{equation}
N_{\mathrm{f},t} \approx N_{\mathrm{f},t-1} \left( 1- p \frac{s M_{\mathrm{i},t-1}}{N_{\mathrm{f},t-1}} \right).
\label{eqn:N_decay}
\end{equation}
By comparing (\ref{eqn:N_decay}) to (\ref{eqn:M}), we can conclude the following.

\begin{Remark}
If the network is overloaded, i.e., if $s M>N$, then $N_{\mathrm{f},t}$ decreases faster than $M_{\mathrm{i},t}$, and all subcarriers tend to be occupied before all nodes have activated. In this case, a large number of contention slots $\kc$ may unnecessarily increase the probability of colliding transmissions, which can affect the energy efficiency by the corresponding alterations in throughput and energy consumption.
\label{remark:decay}
\end{Remark}

Finally, we obtain the energy consumption due to transmissions, as follows.

\begin{Lemma}
The transmission energy consumption $\Et$ per subcarrier in a time slot is given by
\begin{equation}
\Et = \frac{\Pt}{\kf} \left[ \sum_{t=1}^{\kc} \mu_t \, T + \mu_{\kc} \left( \kd\, T - \Ts \right) \right]
\label{eqn:Et}
\end{equation}
with $\mu_t$ given in Corollary \ref{corollary:mu}.
\label{Lemma:Et}
\end{Lemma}
\begin{IEEEproof}
Equation (\ref{eqn:Et}) follows by summing up the transmission energy over the whole frame, by considering that each active node spends a time $\Ts$ for sensing rather than transmitting, and by dividing by the number of slots $\kf$ in a frame.
\end{IEEEproof}

\subsection{Validation and Insights}

We now provide numerical results to confirm the analysis presented in this section and to give insights into the energy consumption incurred at the nodes due to spectrum sensing and transmission under a hybrid signal processing scheme that employs a random access protocol.

\begin{figure}[!t]
\centering
\includegraphics[width=\figwidth]
{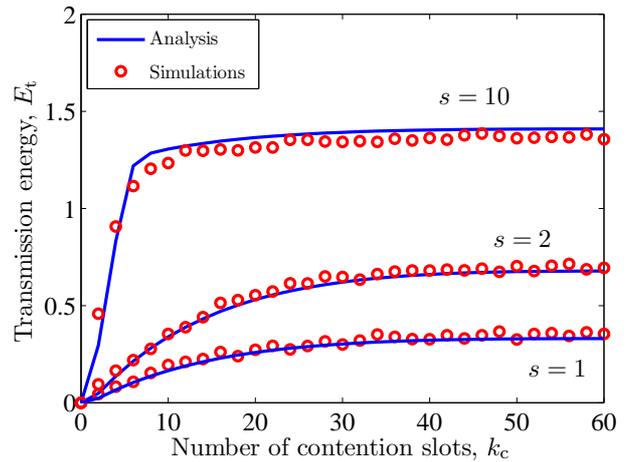}
\caption{Transmission energy consumption $\Et$ versus $\kc$, with $\md=\fa=1\%$, $M=32$ nodes, $N=64$ subcarriers, total frame length of $\kf=60$ slots, $\Pt=1$, $T=1$, $\Ts=0.1$, $p=5\%$, and various values of $s$. }
\label{fig:Et}
\end{figure}

In Fig. \ref{fig:Et}, we compare the transmission energy consumption $\Et$ given in Lemma~\ref{Lemma:Et} to the values obtained from simulations. The value of $\Et$ is plotted versus the number of contention slots $\kc$. Imperfect spectrum sensing is considered, with probabilities of missed detection and false alarm $\md=\fa=1\%$, which are consistent with the values found in Section III. Figure~\ref{fig:Et} shows that simulation results agree well with the analytical values from Lemma~\ref{Lemma:Et}. The figure also shows that by increasing the contention period $\kc$, more subcarriers are likely to be occupied, thus increasing the transmission energy $\Et$. Moreover, Fig.~\ref{fig:Et} confirms the observations made in Remark \ref{remark:decay} by showing that in an overloaded network, i.e., for $s=10$, the transmission energy saturates since the nodes quickly occupy all available subcarriers.

\begin{figure}[!t]
\centering
\includegraphics[width=\figwidth]{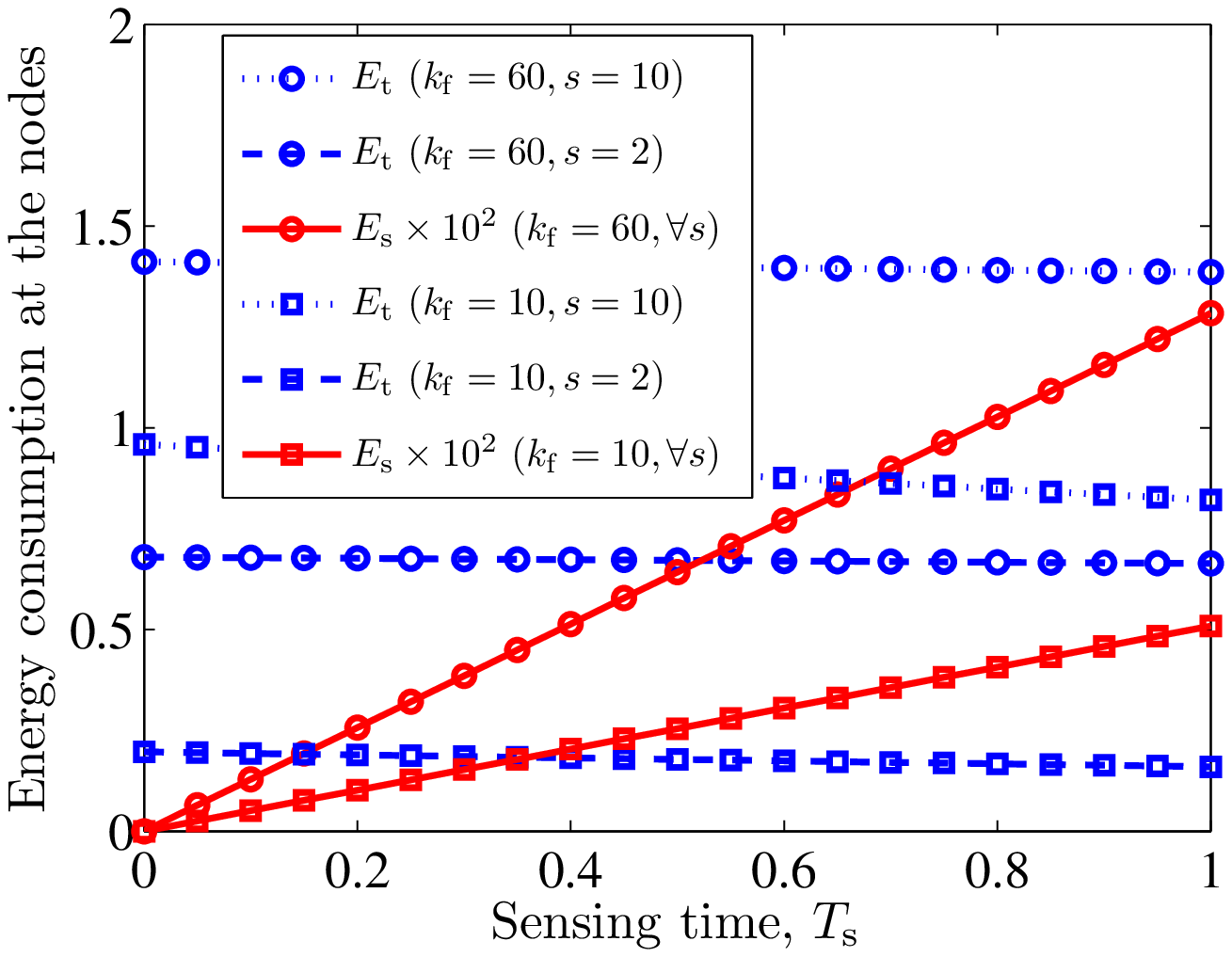}
\caption{Transmission energy $\Et$ and sensing energy consumption $\Es$ (scaled by $\times 10^2$) versus $\Ts$, with $\md=\fa=1\%$, $M=32$ nodes, $N=64$ subcarriers, $\Pt=1$, $\Ps=10^{-2}$, $T=1$, $p=5\%$, and $\kc=\kf$. }
\label{fig:Et_Es}
\end{figure}

In Fig. \ref{fig:Et_Es}, we compare the transmission energy consumption $\Et$ to the sensing energy consumption $\Es$ given in Lemma~\ref{Lemma:Et} and Lemma~\ref{Lemma:Es}, respectively. The values of $\Et$ and $\Es$ are plotted versus the sensing time $\Ts$. Figure \ref{fig:Et_Es} shows that the sensing energy is negligible compared to the transmission energy as long as the sensing time $\Ts$ is small compared to the time slot duration $T$ and the frame duration $\kf$ is long enough. Moreover, the figure shows that the transmission energy is sensitive to the cluster load but not to the sensing time $\Ts$, whereas the opposite is true for the sensing energy.

\section{Multi-User Decoding}

In this section, we analyze the decoding energy consumption when multi-user decoding  is implemented at the AP, and we derive the energy efficiency of a hybrid signal processing scheme.

\subsection{Performance Analysis of Multi-User Decoding}

In a hybrid signal processing scheme, multi-user decoding can be performed at the AP to resolve some of the collisions arising from the combination of an imperfect spectrum sensing scheme and a random access protocol. The idea behind MD is to successively decode multi-user signals according to descending signal power. Therefore, the strongest signal is first decoded and subtracted from the incoming signal, so that interference is reduced, then the second strongest signal is decoded and subtracted, etc. The process is repeated until either all multi-user signals are decoded or decoding fails \cite{Verdu2011}. Multi-user decoding increases the rate but comes at the cost of a higher decoding energy consumption due to the multiple decoding attempts. The performance of MD depends on the order statistics of the received signal power, which in turn is affected by the spatial distribution of the transmitting nodes and on the propagation channel conditions \cite{WilQueKouSlu_SPAWC13, ZhaHae_Asilomar13}. Consistently with previous work \cite{HasAloBasEbb_TWC2003, HuaLauChe_JSAC2009,GeraciDhillonTC}, we consider perfect interference cancellation.

In the following, we explicitly model the sequence of events in the decoding process. We define the success probability as a function of the decoding threshold, the number of decoded transmissions, and all relevant system parameters such as transmission power, path loss exponent, and channel fading. The statistics of the number of colliding transmissions on a given subcarrier at time slot $t$ are determined by the state probabilities $\Pcal_{l,t}$ in (\ref{eqn:P}). Hence, the colliding nodes form a binomial point process (BPP). In this section, we provide analytical results for the probability of successful decoding in the presence of a BPP of colliding nodes. We make the following assumption.

\begin{Assumption}
In our model, we include the effects of both fading and topology, yet, we assume that the order statistics are dominated by the distance. This can be understood by considering that the order statistics of the distance outweigh the fading effects, which vary on a much shorter time scale.
\label{Ass:distance_dom} 
\end{Assumption}
A formal proof for Assumption~\ref{Ass:distance_dom} can be found in \cite{WilQueKouRabSlu_TCOM14}, where it is shown that considering the class of Nakagami-$m$ fading, the order statistics of the received signal power are dominated by the distance. Since the proof in \cite{WilQueKouRabSlu_TCOM14} holds for the tails of the distribution \cite{CliSam_StoPro94}, the accuracy of Assumption~\ref{Ass:distance_dom} will be verified in Fig.~\ref{fig:BPP_vs_PPP}.

Let $l$ be the number of colliding transmissions on a given subcarrier. The decoding order is based on the received signal power. The powers received by the AP from each transmission can be ordered as
\begin{equation}
X_{(1)} \geq X_{(2)} \geq \ldots \geq X_{(l)}
\end{equation}
where
\begin{equation}
X_{(n)} = \Pt |h_n|^2 D_{(n)}^{-\alpha}	
\end{equation}
is the power received from the $n$-th strongest node, and $h_n$ and $D_{(n)}$ are the fading coefficient and the distance between the $n$-th strongest node and the AP, respectively.
By assuming the noise negligible compared to the interference, we have that the decoding of the $n$-th strongest transmission is successful if
\begin{align} 
\frac{X_{(n)}}{I_{\Omega_n} + \sigma^2_{\mathrm{I}}} \geq \zeta \quad \forall n \leq l
\end{align}
where $\zeta$ is the decoding threshold, $\sigma^2_{\mathrm{I}}$ is the interference originating from other clusters given in (\ref{eqn:moments2}), and $I_{\Omega_n}$ represents the aggregate interference originating from the representative cluster after canceling $n$ transmissions, given by
\begin{equation}
I_{\Omega_n} = \sum_{i = n+1}^{l} X_{(i)}.
\label{eq.aggIntSIC}
\end{equation}

We now give the following result on the probability of successfully decoding the $n$-th strongest transmission.

\begin{Lemma}
The probability $\Pdecl(n)$ of successfully decoding the $n$-th strongest transmission given the correct decoding of the $n-1$ strongest transmissions, under $l \geq n$ colliding transmissions, is given by
\begin{equation}
\Pdecl(n) = \int_0^{\rc} \Pdecl(n \,|\, x) f_{D_{(n)}}(x) \mathrm{d}x
\label{eqn:Pdecl}
\end{equation}
with
\begin{align}
&\Pdecl(n \,|\, x) = \exp(-\zeta x^\alpha \sigma^2_{\mathrm{I}})\nonumber\\
\!&\times \! \left( \frac{1}{\rc^2\!-\!x^2} \left( y \!-\! y \,_2F_1\left(1, \frac{2}{\alpha}, 1 \!+\! \frac{2}{\alpha}, - \frac{x^\alpha y^{\alpha/2}}{\zeta}\right) \right) \right)^{l-n}
\label{eqn:Pdecl_cond}
\end{align}
and
\be
f_{D_{(n)}}(x) = \frac{1}{\mathcal{B}(n, l-n+1)} F_D^{n-1}(x)[1-F_D(x)]^{l-n} f_D(x),
\ee
and where $f_D(x) = 2x/\rc^2$, $F_D(x) = x^2/\rc^2$, and we denoted by $_2F_1(.)$ the Gaussian hypergeometric function and by $\mathcal{B}(a, b) = \int_0^1 t^{a-1} (1-t)^{b-1} \mathrm{d}t$, $a>0$, $b>0$, the beta function. 
\label{Lemma:Pdec}
\end{Lemma}
\begin{IEEEproof}
See Appendix \ref{App:Pdec}.
\end{IEEEproof}

\begin{Corollary}
For path loss exponent $\alpha = 4$, the conditional probability $\Pdecl(n|x)$ in (\ref{eqn:Pdecl_cond}) reduces to
\begin{align}
\! & \Pdecl(n | x) = \exp(-\zeta x^4 \sigma^2_{\mathrm{I}})\nonumber\\
&\! \times \Bigg[\!1 \!+\! \frac{\sqrt{\zeta} x^2}{\rc^2\!-\!x^2} \! \left(\! \tan^{-1}\!\left( \frac{1}{\sqrt{\zeta}} \right) \!\!-\! \tan^{-1}\!\left(\frac{\rc^2}{\sqrt{\zeta}x^2} \right) \! \right) \! \Bigg]^{l-n} \,\,.
\end{align}
\end{Corollary}
\begin{IEEEproof}
The corollary follows by noting that for $\alpha = 4$ the integral $I_n(\alpha)$ in (\ref{eqn:In}) reduces to
\be
I_n(4) = y - \sqrt{\zeta} x^2 \tan^{-1} \left( \frac{y}{\sqrt{\zeta}x^2} \right).
\ee
\end{IEEEproof}

The statistics of the number of successfully decoded transmissions can now be obtained as follows.

\begin{Lemma}
The probability $\Dcal_{i,l}$ of correctly decoding $i$ out of $l$ colliding transmissions is given by
\begin{equation}
\Dcal_{i,l} = \left[ 1 - \Pdecl(i+1) \right] \cdot \prod_{n=1}^{i} \Pdecl(n)	
\label{eq.Dil}
\end{equation}
with $\Pdecl(n)$ given in (\ref{eqn:Pdecl}) if $l \geq n$, and $\Pdecl(n)=0$ otherwise.
\label{Lemma:Dil}
\end{Lemma}
\begin{IEEEproof}
The AP successfully decodes $i$ colliding transmissions if
\begin{align} 
\frac{X_{(n)}}{I_{\Omega_n} + \sigma^2_{\mathrm{I}}} \geq \zeta \enspace \forall n \leq i \enspace \textrm{and} \enspace \frac{X_{(i+1)}}{I_{\Omega_{i+1}} + \sigma^2_{\mathrm{I}}} < \zeta.
\end{align}
The lemma then follows by assuming the independence between the consecutive decoding of transmissions \cite{WilQueKouRabSlu_TCOM14}.
\end{IEEEproof}

\subsection{Energy Efficiency}

We can now obtain the mean energy consumption due to decoding at the AP, as follows.

\begin{Lemma}
The decoding energy consumption $\Ed$ per subcarrier per time slot is given by
\begin{align}
\Ed &= \frac{\Pd \, T}{\kf} \left[\sum_{t=1}^{\kc} \sum_{l=1}^{M} \Pcal_{l,t} \sum_{i=0}^{l} (i+1) \Dcal_{i,l} \right. \nonumber\\
& \left. \quad + \kd \sum_{l=1}^{M} \Pcal_{l,\kc} \sum_{i=0}^l (i+1) \Dcal_{i,l}\right].
\label{eqn:Ed}
\end{align}
\label{Lemma:Ed}
\end{Lemma}
\begin{IEEEproof}
Equation (\ref{eqn:Ed}) follows since the decoding energy consumption is proportional to the number of decoding attempts performed at the AP, i.e., the number of successfully decoded transmissions plus one, and by using Lemma \ref{Lemma:Dil} and dividing by the number of slots $\kf$ in a frame.
\end{IEEEproof}

We now derive the throughput of a hybrid signal processing scheme, i.e., the mean number of bits successfully transmitted on each subcarrier per time slot.

\begin{Lemma}
The throughput $R$ of a hybrid signal processing scheme with spectrum sensing, media access control, and multi-user decoding is given by
\begin{align}
R = \frac{\chi(\zeta) \, T}{\kf} \left[\sum_{t=1}^{\kc} \sum_{l=1}^{M} \Pcal_{l,t} \sum_{i=1}^{l} i \Dcal_{i,l} + \kd \sum_{l=1}^{M} \Pcal_{l,\kc} \sum_{i=1}^l i \Dcal_{i,l} 
 \right]
\label{eqn:R}
\end{align}
where the probabilities $\Dcal_{i,l}$ and $\Pcal_{l,t}$ are given in (\ref{eq.Dil}) and (\ref{eqn:P}), respectively.
\label{Lemma:R}
\end{Lemma}
\begin{IEEEproof}
The lemma follows by calculating the average number of successfully decoded transmissions during the contention period and the contention-free period, respectively, by neglecting the small amount of time spent for spectrum sensing, and by dividing by the number of slots $\kf$ in a frame.
\end{IEEEproof}

We finally obtain the energy efficiency $\eta$, defined as the number of bits successfully transmitted per joule of energy spent.

\begin{Theorem}
The energy efficiency $\eta$ under a hybrid signal processing scheme with spectrum sensing, media access control, and multi-user decoding is given by
\be
\eta=\frac{R}{\Es+\Et+\Ed} \, \left[\frac{\mathrm{bits}}{\mathrm{J}}\right],
\ee
\label{eqn:eta}
where $R$, $\Es$, $\Et$, and $\Ed$ are given by (\ref{eqn:R}), (\ref{eqn:Es}), (\ref{eqn:Et}), and (\ref{eqn:Ed}), respectively.
\label{theorem:eta}
\end{Theorem}
\begin{IEEEproof}
The theorem follows from Lemma \ref{Lemma:Es}, Lemma \ref{Lemma:Et}, Lemma \ref{Lemma:Ed}, and Lemma \ref{Lemma:R}, and by dividing the throughput by the whole energy consumption incurred in each cluster, i.e., both at the nodes and at the AP, in one time slot.
\end{IEEEproof}

\subsection{Validation and Insights}

\begin{figure}[!t]
\centering
\includegraphics[width=\figwidth]{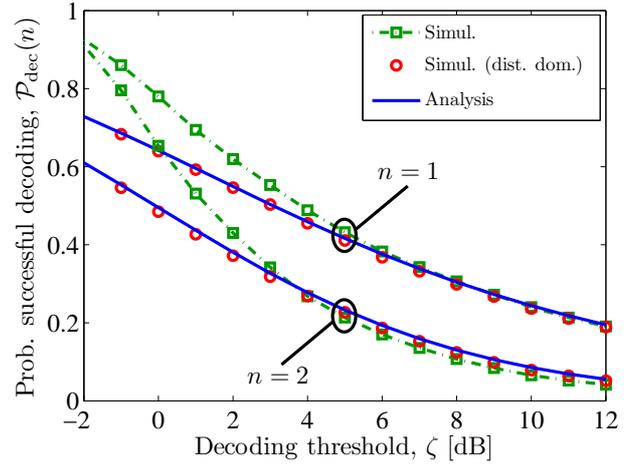}
\caption{Probability of successful decoding as a function of the decoding threshold $\zeta$ for the strongest and second strongest colliding transmissions, with cluster radius $\rc = 100$ and $l = 5$ colliding nodes. Analytical values from Lemma \ref{Lemma:Pdec} are compared to simulations.}
\label{fig:BPP_vs_PPP}
\end{figure}

In Fig. \ref{fig:BPP_vs_PPP}, we compare the simulated probability of successful decoding to simulations obtained in the case of distance-dominated order statistics and to analytical results from Lemma \ref{Lemma:Pdec}. The figure shows that for practical values of the decoding threshold $\zeta$, the probability of successful decoding can be well approximated by assuming that the order statistics are dominated by the distance, thus justifying Assumption \ref{Ass:distance_dom}. Under distance-dominated order statistics, Fig. \ref{fig:BPP_vs_PPP} confirms also the accuracy of the analysis in Lemma \ref{Lemma:Pdec} for all values of the threshold $\zeta$.

\begin{figure}[!t]
\centering
\includegraphics[width=\figwidth]{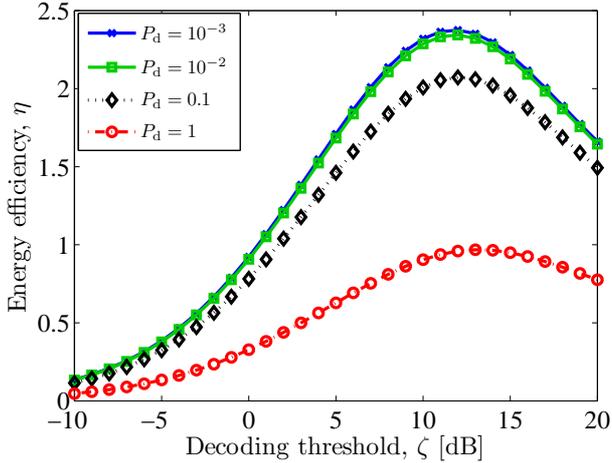}
\caption{Energy efficiency $\eta$ as a function of the decoding threshold $\zeta$ with $\md=\fa=1\%$, $M=32$ nodes, $N=64$ subcarriers, $s=3$, $\rc = 100$, $\kf=60$, $p=5\%$, $\Pt=1$, $\Ps=10^{-2}$, $T=1$, $\Ts=0.1$, and $\chi=\log_2(1+\zeta)$.}
\label{fig:eta_vs_zeta}
\end{figure}

Figure \ref{fig:eta_vs_zeta} shows the energy efficiency of a hybrid signal processing scheme as a function of the decoding threshold $\zeta$. The figure shows that the energy efficiency does not have a monotonic behavior, since it depends on a tradeoff between the probability of successful decoding, which decreases with $\zeta$, and the spectral gain $\chi(\zeta)$, which increases with $\zeta$. Although Fig. \ref{fig:eta_vs_zeta} shows that the maximum value of $\eta$ is achieved for $\zeta \approx 12$dB, we note from Fig. \ref{fig:BPP_vs_PPP} that, depending on the number of collisions, this may correspond to a case when almost none of the colliding transmission can be decoded. This case could be undesirable, and in practice one may design the system to work at lower values of $\zeta$ and impose a constraint on the success probability $\Pdecl(n)$.

\section{Energy Efficiency Tradeoff}

In this section, we compare the energy efficiency of a hybrid signal processing scheme to the one obtained with centralized and distributed approaches. In the following, we use the subscripts $\mathrm{C}$ and $\mathrm{D}$ to denote energy consumption, throughput, and energy efficiency under centralized and distributed signal processing, respectively.

\subsection{Centralized Signal Processing}

When the signal processing operations are performed in a centralized fashion, the AP schedules the transmission of all nodes by means of a polling mechanism performed via a control channel. In this case, neither a spectrum sensing scheme nor a random access MAC protocol are necessary at the nodes. Moreover, a multi-user decoding scheme is not needed either since collisions are avoided by the centralized access scheme. We denote by $P_{\mathrm{c}}$ the power consumption due to the use of a control channel, and by $E_{\mathrm{C,c}}=P_{\mathrm{c}} T$ the energy consumption on the control channel per subcarrier and per time slot.

\begin{Lemma}
The energy consumption $E_{\mathrm{C}}$ per subcarrier in a time slot under a centralized signal processing scheme is given by
\begin{equation}
E_{\mathrm{C}} = \left( \Pc + \Pt + \Pd \right) T.
\label{eqn:EC}
\end{equation}
\label{Lemma:EC}
\end{Lemma}
\begin{IEEEproof}
Under a centralized signal processing scheme, the transmission and decoding energy are given by $E_{\mathrm{C,t}}=\Pt T$ and $E_{\mathrm{C,d}}=\Pd T$, respectively, since we have only one transmission per subcarrier and the AP needs to perform one decoding attempt only. Adding the energy consumption $E_{\mathrm{C,c}}=P_{\mathrm{c}} T$ due to the control channel yields (\ref{eqn:EC}).
\end{IEEEproof}

\begin{Theorem}
The energy efficiency $\eta_{\mathrm{C}}$ of a centralized signal processing scheme is given by
\begin{align}
\eta_{\mathrm{C}} = \frac{\chi(\zeta) \, \Dcal_{1,1}}{\Pc + \Pt + \Pd} \, \left[\frac{\mathrm{bits}}{\mathrm{J}}\right].
\label{eqn:eta_C}
\end{align}
\label{theorem:eta_C}
\end{Theorem}
\begin{IEEEproof}
The theorem follows from Lemma \ref{Lemma:EC} and by considering that the throughput under centralized signal processing is given by $R_{\mathrm{C}}=\chi(\zeta) \Dcal_{1,1} T$.
\end{IEEEproof}

\subsection{Distributed Signal Processing}

When the signal processing operations are performed in a distributed way, access to the spectrum is obtained through a generic MAC protocol that builds on a spectrum sensing functionality at each node. In this case, no control channel, scheduling, and multi-user decoding are needed at the AP, which simply decodes one single transmission per subcarrier. 

\begin{Lemma}
The energy consumption $E_{\mathrm{D}}$ per subcarrier in a time slot under a distributed signal processing scheme is given by
\begin{equation}
E_{\mathrm{D}} = \Es + \Et + \Pd T
\label{eqn:ED}
\end{equation}
with $\Es$ and $\Et$ given in (\ref{eqn:Es}) and (\ref{eqn:Et}), respectively.
\label{Lemma:ED}
\end{Lemma}
\begin{IEEEproof}
Under a distributed signal processing scheme, the sensing and transmission energies are the same as the ones derived for the hybrid signal processing scheme in Lemma~\ref{Lemma:Es} and Lemma~\ref{Lemma:Et}, respectively. Moreover, since no multi-user decoding is implemented at the AP, only one decoding attempt is required, and the decoding energy is given by $E_{\mathrm{D,d}} = \Pd T$.
\end{IEEEproof}

\begin{Lemma}
The throughput $R_{\mathrm{D}}$ of a distributed signal processing scheme is given by
\begin{align}
R_{\mathrm{D}} = \frac{\chi(\zeta) \, T}{\kf} \left\{\sum_{t=1}^{\kc} \sum_{l=1}^{M} \Pcal_{l,t} \Dcal_{1,l} + \kd \sum_{l=1}^{M} \Pcal_{l,\kc} \Dcal_{1,l} \right\}
\label{eqn:RD}
\end{align}
where the probabilities $\Dcal_{1,l}$ and $\Pcal_{l,t}$ are given in (\ref{eq.Dil}) and (\ref{eqn:P}), respectively.
\label{Lemma:RD} 
\end{Lemma}
\begin{IEEEproof}
The lemma follows by (i) noting that in the absence of a multi-user decoding scheme only one decoding attempt is performed at the AP, (ii) calculating the average number of successful single decoding attempts during the contention period and the contention-free period, respectively, and (iii) neglecting the small amount of time spent for spectrum sensing.\end{IEEEproof}

\begin{Theorem}
The energy efficiency $\eta_{\mathrm{D}}$ of a distributed signal processing scheme is given by
\be
\eta_{\mathrm{D}} = \frac{R_{\mathrm{D}}}{E_{\mathrm{D}}} \, \left[\frac{\mathrm{bits}}{\mathrm{J}}\right].
\label{eqn:eta_D}
\ee
with $R_{\mathrm{D}}$ and $E_{\mathrm{D}}$ given in (\ref{eqn:RD}) and (\ref{eqn:ED}), respectively.
\label{theorem:eta_D}
\end{Theorem}
\begin{IEEEproof}
The theorem follows from Lemma \ref{Lemma:Es}, Lemma \ref{Lemma:Et}, Lemma \ref{Lemma:ED}, and Lemma \ref{Lemma:RD}.
\end{IEEEproof}

\subsection{Numerical Results}

\begin{figure}[!t]
\centering
\includegraphics[width=\figwidth]{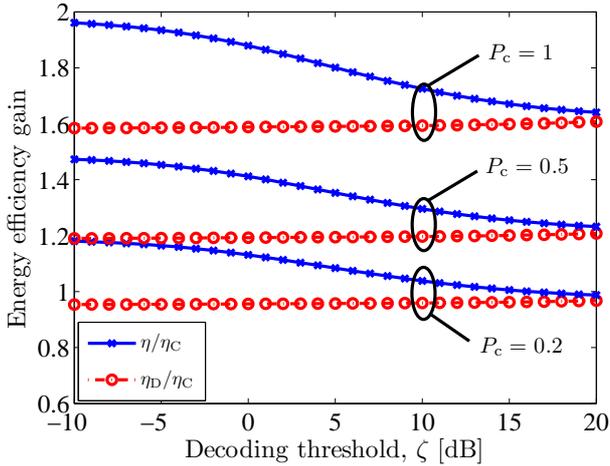}
\caption{Energy efficiency of hybrid and distributed signal processing normalized to the one of a centralized scheme versus $\zeta$, with $\md=\fa=1\%$, $M=32$ nodes, $N=64$ subcarriers, $s=3$, $\rc = 100$, $\kf=60$, $p=5\%$, $\Pt=1$, $\Ps=\Pd=10^{-2}$, $T=1$, $\Ts=0.1$, and $\chi=\log_2(1+\zeta)$.}
\label{fig:tradeoff_vs_zeta}
\end{figure}

Figure \ref{fig:tradeoff_vs_zeta} shows the energy efficiency of hybrid and distributed signal processing normalized by the one of a centralized scheme versus the decoding threshold $\zeta$. The figure shows that hybrid or distributed approaches can be preferable to a centralized approach, especially when the power $\Pc$ consumed on the control channel is comparable to the transmit power $\Pt$. Moreover, Fig. \ref{fig:tradeoff_vs_zeta} shows that a hybrid signal processing scheme outperforms a distributed approach, especially for lower values of $\zeta$ and $\Pd$, since multi-user decoding is successful and it does not incur a high energy consumption. Finally, the figure shows that the two curves converge for higher values of $\zeta$, when it is not worth attempting to decode more than one transmission.

\begin{figure}[!t]
\centering
\includegraphics[width=\figwidth]{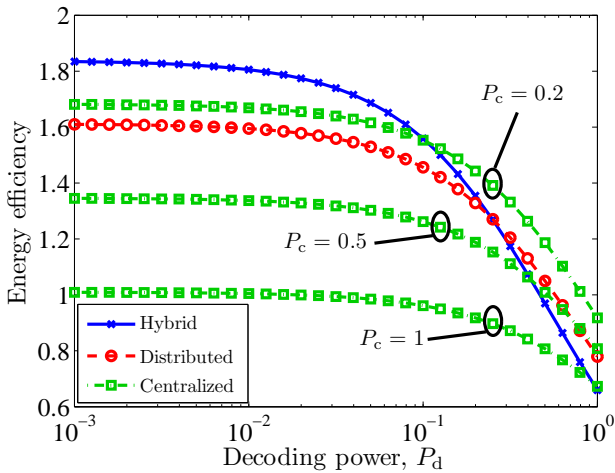}
\caption{Energy efficiency of hybrid, distributed, and centralized signal processing versus decoding power $\Pd$, with $\md=\fa=1\%$, $M=32$ nodes, $N=64$ subcarriers, $s=3$, $\rc = 100$, $\kf=60$, $p=5\%$, $\Pt=1$, $\Ps=10^{-2}$, $T=1$, $\Ts=0.1$, $\zeta=5$dB, and $\chi=\log_2(1+\zeta)$.}
\label{fig:tradeoff_vs_Pd}
\end{figure}

Figure \ref{fig:tradeoff_vs_Pd} compares the energy efficiency of hybrid signal processing to the one obtained with distributed and centralized approaches as a function of the decoding power $\Pd$. The energy efficiency of the centralized scheme is affected by the power consumption of the control channel, $\Pc$. The figure shows that for relatively high values of the decoding power, i.e., $\Pd\geq\Pt/3$, it is preferable to employ either a centralized or a distributed approach (depending on the control channel power consumption $\Pc$). However, for lower and more practical values of the decoding power, hybrid signal processing outperforms both the distributed and the centralized approaches.

\section{Conclusion}

In this work, we introduced an analytical toolset to evaluate the energy efficiency of wireless networks under centralized, distributed, and hybrid signal processing. For the three scenarios above, we proposed a cross-layer approach to derive the throughput and the energy consumption due to signal processing operations. We used a general model that accounts for the clustered nature of wireless networks, for a practical MAC protocol that relies on an imperfect spectrum sensing functionality, and for decoding operations in the presence of interference. Our results revealed that in practical cases a hybrid approach, where the signal processing operations are shared between nodes and APs, can achieve energy efficiency gains over fully centralized or fully distributed schemes.

The framework provided in this paper allows for a jointly optimal design of the MAC and physical layers, since it gives insights into how the energy efficiency is affected by several system parameters, i.e., sensing time, spectrum access probability, density of nodes, size of clusters, frame length, etc. More generally, our work helps to understand how the distribution of the signal processing load, which is essential for the feasibility of future dense wireless networks, will affect the energy efficiency. A lower energy consumption may also help reduce the complexity of the devices that need to be deployed, i.e., APs may become smaller and require less cooling. Investigating the cascading effect of the signal processing load on these additional contributions to energy consumption is regarded as an interesting future research direction.
\appendices

\section{}
\begin{IEEEproof}[Proof of Lemma \ref{Lemma:moments}] \label{App:moments}
To characterize the interference distribution, the parameters of the Gaussian distribution are found by matching the cumulants of the aggregate interference with the cumulants of the Gaussian distribution. Based on \cite{WinPinShe_ProcIEEE2009, RabQueShiWin_JSAC2011} and using the PPP assumption for the active node set, the characteristic function (CF) of the aggregate interference can be expressed as
\begin{align} \nonumber
\psi_{\Ibk}(j\omega)
=&\exp\!\Bigg(\!\!-2\pi \mu \lambda_\mathrm{h} \! \int_0^\infty \! \int_{d_{\mathrm{min}}}^\infty \!\!\left[ 1 \!- \!\exp(j \omega \sqrt{\Pt} g r^{\alpha/2})\right] \\ 
&\times f_{|h|}(g) r \mathrm{d}r \mathrm{d}g\Bigg)  .
\end{align}
Consequently, the cumulants of $\Ibk$ can be written as 
\begin{align} \nonumber
\kappa_{\Ibk}(n) &= \frac{1}{j^n}\frac{\partial^n \ln \psi_{\Ibk}(j\omega)}{\partial \omega^n} \Bigg|_{\omega = 0} \\
& = \Pt^{n/2} \frac{\pi \mu \lambda_\mathrm{h}}{n\alpha -1} d_\mathrm{min}^{2-n\alpha/2} \mu_{|h|,n}\, ,
\end{align}
where $\mu_{|h|,n}$ represents the $n$-th moment of the fading distribution. The CF of the Gaussian distribution is given by 
\be
\psi_\mathcal{N}(j\omega) = \exp \left(\mu_{\mathrm{I}} j\omega - \sigma_{\mathrm{I}}^2 \omega^2 \right)\, .
\ee
By matching the first two cumulants of the aggregate interference and the corresponding Gaussian distribution, we obtain the moments in Lemma \ref{Lemma:moments}.
\end{IEEEproof}

\section{}
\begin{IEEEproof}[Proof of Lemma \ref{Lemma:PmdPfa}] \label{App:PmdPfa}
In order to calculate $\md$, we use a generic approach \cite{WilQueRabSlu_TC2013} based on the characteristic function (CF) of the decision variable for a typical user, i.e., a user with random location within the cluster. Let $\tildeWbk \triangleq \frac{\Wbk}{\sqrt{B}}$ with $\tildeWbk \sim \mathcal{CN}(0, \sigma^2_{\tilde{\mathrm{W}}})$ and $\sigma_{\tilde{\mathrm{W}}}^2 \triangleq \frac{\sigma_{\mathrm{w}}^2}{B}$, and let $\tildeIbk \triangleq \frac{\Ibk}{\sqrt{B}}$ with $\tildeIbk \sim \mathcal{CN}(0, \sigma^2_{\tilde{\mathrm{I}}})$ and $\sigma_{\tilde{\mathrm{I}}}^2 \triangleq \frac{\sigma_{\mathrm{I}}^2}{B}$. Under hypothesis $\mathcal{H}_1$, the decision variable of the energy detector in (\ref{eqn:Ek}) can be rewritten as
\begin{equation}
E_k = \sum_{b = 1}^{B}\Big(\tildeSbk + \tildeIbk + \tildeWbk \Big)^2.
\end{equation}
Both the interference term and the noise term are Gaussian r.v.'s that can be merged as follows
\begin{equation}
\tildeNbk \triangleq \frac{\Ibk+\Wbk}{\sqrt{B}} \, ,
\end{equation}
where $\tildeNbk \sim \mathcal{CN}(0, \sigma^2_\mathrm{IN})$ and $\sigma^2_\mathrm{IN} = \sigma^2_{\tilde{\mathrm{I}}} + \sigma^2_{\tilde{\mathrm{W}}}$. To define the CF of the decision variable $\psi_{E_k|\Hone}(j\omega) = \mathbb{E}[\exp(j\omega E_k)]$, we first condition on $\tildeSbk$, such that $E_k$ follows a non-central chi-square distribution with CF given by
\begin{equation}
\psi_{{E_k}|\Hone,\tildeSbk}(j\omega) = \frac{1}{\left( 1-2j\omega\sigma_{\mathrm{IN}}^2 \right)^{B/2}} \exp\left(\frac{j\omega B \tildeSbk^2}{ 1-2j\omega\sigma_{\mathrm{IN}}^2} \right).
\label{eq.CFdecVarConT}
\end{equation}
Let $\mathcal{J}$ be the set of active nodes on subcarrier $k$ in the representative cluster, with $l= |\mathcal{J}|$. If we assume that all the signals transmitted by nodes $j \in \mathcal{J}$ have a normal distribution, then $\tildeSbk^2$ represents the power of a normally distributed r.v. with variance $\sum_{i \in \mathcal{J}} |h_i|^2 \Pt/(B d_i^{\alpha})$. We can then obtain the CF by taking the expectation over the distance and fading parameters of the signal of interest, which can be written as follows
\begin{align}
\psi_{E_k|\Hone}(j\omega) &= \frac{\mathbb{E}_{h_i,d_i}\left[\exp\left(\frac{\sum_{i=1}^l j\omega \Pt |h_i|^2/d_i^\alpha}{1 - 2 j\omega\sigma_\mathrm{IN}^2}\right) \right]}{\left( 1-2j\omega\sigma_{\mathrm{IN}}^2 \right)^{B/2}} \\ \nonumber
&= \frac{\mathbb{E}_{h_i,d_i}\left[ \prod_{i=1}^{l}\exp\left(\frac{j\omega \Pt |h_i|^2/d_i^\alpha}{1 - 2 j\omega\sigma_\mathrm{IN}^2}\right) \right]}{\left( 1-2j\omega\sigma_{\mathrm{IN}}^2 \right)^{B/2}}.
\end{align}
As the variables $d_i$ and $h_i$ are all independent, the expectation can be brought inside the product and we find
\begin{align}
\psi_{E_k|\Hone}(j\omega) &= \frac{\prod_{i=1}^{l}  \mathbb{E}_{h_i,d_i} \left[ \exp\left(\frac{j\omega \Pt |h_i|^2/d_i^\alpha}{1 - 2 j\omega\sigma_\mathrm{IN}^2}\right) \right]}{\left( 1-2j\omega\sigma_{\mathrm{IN}}^2 \right)^{B/2}} \nonumber \\
&= \frac{\left(  \mathbb{E}_{d_i} \left[ \frac{1}{1 - j\omega(\Pt/(2 d_i^\alpha) + \sigma_\mathrm{IN}^2} \right] \right)^l
}{\left( 1-2j\omega\sigma_{\mathrm{IN}}^2 \right)^{B/2-1}} 
\end{align}
where $d_i$ is the distance between the typical sensing node and any other node $i$ in the same cluster, with pdf as in (\ref{eqn:pdf_ri}). Equation (\ref{eqn:md}) follows by solving the expectation and by using the inversion theorem. 

For the calculation of $\fa$, we apply the same methodology but in the absence of the signal of interest $\Sbk$. The CF of the decision variable $E_k$ under hypothesis $\Hzero$ can be expressed as
\be
\psi_{{E_k}|\mathcal{H}_0}(j\omega) = \frac{1}{\left( 1-2j\omega\sigma_{\mathrm{IN}}^2 \right)^{B/2}}
\ee 
and $\fa$ can be obtained by applying the inversion theorem.
\end{IEEEproof}

\section{}
\begin{IEEEproof}[Proof of Lemma \ref{Lemma:P}] \label{App:P}
Equation (\ref{eqn:Til}) follows by conditioning on a given subcarrier being occupied by $i>0$ nodes at the beginning of time slot $t$, and by obtaining the probability that $l-i$ nodes start new transmissions on the subcarrier during time slot $t$ due to missed detection events. 
Equation (\ref{eqn:T0l}) follows from the probability $\Acal_{i,t}$ that $i$ nodes activate at time slot $t$ and the probability $\Gcal_{i,l,t}$ that $l$ nodes choose the subcarrier given that $i$ nodes have activated.
Equation (\ref{eqn:T00}) follows from the probability that none of the inactive nodes activate and choose the subcarrier at time slot $t$. The probabilities $\mathcal{P}_{l,t}$ can then be obtained recursively since the states $\mathbf{S}_l$ form a Markov chain with transition probabilities $\Tcal_{i,l,t}$.
\end{IEEEproof}

\section{}
\begin{IEEEproof}[Proof of Lemma \ref{Lemma:Pdec}] \label{App:Pdec}
In order to obtain the probabilities $\Pdecl(n)$, $n=1,\ldots,l$, we first obtain $\Pdecl(n|D_{(n)})$, $n=1,\ldots,l$ by conditioning on the distance between the AP and the $n$-th stronger node, and by using a BPP $\Omega_n$ that denotes the representative cluster after $n$ cancellations. We have
\begin{align}
&\Pdecl(n \,|\, D_{(n)}) = \Pr\left[\frac{X_{(n)}}{I_{\Omega_n} + \sigma^2_{\mathrm{I}}} \geq \zeta | D_{(n)} \right] \nonumber\\
& \enspace = \Pr \left[|h_n|^2 \geq \zeta D_{(n)}^\alpha (I_{\Omega_n} + \sigma^2_{\mathrm{I}}) \, | \, D_{(n)}\right]  \nonumber \\
& \enspace = \mathbb{E}_{\Omega_n, |h|^2} \left[ \exp(-\zeta D_{(n)}^\alpha I_{\Omega_n}) \right] \exp(-\zeta D_{(n)}^\alpha \sigma^2_{\mathrm{I}}) \nonumber \\
& \enspace \overset{(a)}{=} \mathbb{E}_{\Omega_n} \left[ \prod_{x \in \Omega_n} \mathbb{E}_{|h|^2} \Big\lbrace \exp(-\zeta D_{(n)}^\alpha |h_x|^2 x^{-\alpha}) \Big\rbrace \right] e^{-\zeta D_{(n)}^\alpha \sigma^2_{\mathrm{I}}}\nonumber\\
& \enspace = \mathbb{E}_{\Omega_n} \left[ \prod_{x \in \Omega_n} \frac{1}{1+\zeta\left(\frac{D_{(n)}}{x}\right)^\alpha} \right] e^{-\zeta D_{(n)}^\alpha \sigma^2_{\mathrm{I}}}\nonumber\\
& \enspace \overset{(b)}{=} \left( \frac{2 \pi}{\pi (\rc^2 - D_{(n)}^2)} \int_{D_{(n)}}^{\rc} \frac{x}{1+\zeta \left(\frac{D_{(n)}}{x}\right)^\alpha}  \mathrm{d}x \right)^{l-n} \!\!\!\!\! \cdot e^{-\zeta D_{(n)}^\alpha \sigma^2_{\mathrm{I}}} \nonumber \\
& \enspace =  \left( \frac{1}{\rc^2-D_{(n)}^2} \underbrace{\int_{D_{(n)}^2}^{\rc^2} \frac{y^{\alpha/2}}{y^{\alpha/2}+\zeta D_{(n)}^\alpha}   \mathrm{d}y}_{I_n(\alpha)}\right)^{l-n} \!\!\!\!\! \cdot e^{-\zeta D_{(n)}^\alpha \sigma^2_{\mathrm{I}}}
\label{eqn:P_dec_l_cond}
\end{align}
where $(a)$ is due to the independence of the fading parameters, and $(b)$ follows from the probability generating functional (PGFL) of a BPP and since the order statistics are dominated by the distance. The indefinite integral $I_n(\alpha)$ can be solved as
\be
I_n(\alpha) = y - y \,_2F_1(1, 2/\alpha, 1 + 2/\alpha, - D_{(n)}^\alpha y^{\alpha/2} / \zeta).
\label{eqn:In}
\ee
The distribution of the distance $D$ between a random node in the circular cluster of radius $\rc$ and the AP is given by $f_D(x) = 2x/\rc^2$ and $F_D(x) = x^2/\rc^2$. Given $l$ colliding nodes in the cluster, the distribution of the distance $D_{(n)}$ between the AP and the $n$-th strongest node is given by \cite{DavNag_OrdSta2003}
\be \label{eq.distDistBPP}
f_{D_{(n)}}(x) = \frac{1}{\mathcal{B}(n, l-n+1)} F_D^{n-1}(x)[1-F_D(x)]^{l-n} f_D(x) .
\ee
Deconditioning over $D_{(n)}$, we obtain (\ref{eqn:Pdecl}).
\end{IEEEproof}
\ifCLASSOPTIONcaptionsoff
  \newpage
\fi
\balance
\bibliographystyle{IEEEtran}
\bibliography{Bib_Giovanni,StringDefinitions,femtocellsAndInterference,MAC_SIC,multiTierInterference}

\end{document}